# Scale Transfer in 1849 : Heinrich Schwabe to Rudolf Wolf
*Solar Physics*

**Bhattacharya S.**[1,2] · **Lefèvre L.**[1] · **Hayakawa H.**[3] · **Jansen M.**[2] · **Clette F.**[1]

© Springer ••••

**Abstract**
The focus of this study is to reveal the reason behind a scale problem detected around 1849 in the historical version of the International Sunspot Number Series, i.e. version 1 (Leussu et al, Astronomy and Astrophysics, 559, A28, 2013; Friedli, Solar Phys.291, 2505, 2016). From 1826 to 1848 Heinrich Schwabe's observations were considered primary by Rudolf Wolf, and a shift of primary observer from Schwabe to Wolf in 1849 seems to have led to an inconsistency in the Sunspot Number series. In this study we benefited from various datasets, the most important being Schwabe's raw counts from the *Mittheilungen* (Prof. Wolf's Journals) that have been digitised at the Royal Observatory of Belgium between 2017 and 2019. We provide a robust quantification of the detected problem by using classic algebraic calculations but also different methods such as a method inspired by Lockwood et al (Journal of Geophysical Research (Space Physics), 119(7), 5172, 2014), hence assigning a modern k-factor to Schwabe's observations before 1849. We also assess the implications of this 1849 inconsistency on the International Sunspot Number series (Versions 1 and 2) before and after 1849.

✉ B. S.
shreya.bhattacharya@oma.be

L. L.
laure.lefevre@oma.be

H. H.
hisashi@nagoya-u.jp

J. M.
maarten.jansen@ulb.ac.be

C. F.
frederic.clette@oma.be

[1] Royal Observatory of Belgium, WDC-SILSO

[3] Nagoya University, Japan

[2] Université Libre de Bruxelles, Bruxelles, Belgium

## 1. Introduction

Visual sunspot observations form the longest scientific record of solar activity, spanning over four centuries. Long term solar studies are crucial to predict the future evolution of the solar cycles and to improve our understanding of the solar influence on the Earth's climate. A collection of stable visual observations of sunspots and a methodology to stitch them together in a single series is essential to reconstruct the historical solar activity (Vaquero *et al.*, 2016).

International collaborations are active throughout the Solar-Physics community to recalibrate series for the sunspot number (SN) (Clette *et al.*, 2016) and the group number (GN) with further archival investigations and sophisticated calibration methods (Usoskin *et al.*, 2016; Svalgaard and Schatten, 2016; Chatzistergos *et al.*, 2017; Carrasco *et al.*, 2020; Hayakawa *et al.*, 2020; Usoskin, Kovaltsov, and Kiviaho, 2021). Gathering data from all observers, assessing the scale transfers between them and their homogeneity is important for the calibration of the sunspot series. The WDC-SILSO team has been working on this issue through the VAL-U-SUN (http://www.sidc.be/valusun/) Belgian BRAIN project, the organization of sunspot workshops (https://ssnworkshop.fandom.com/wiki/Home), editorial work for a Solar Physics topical issue on SN re-calibration (Friedli, 2016) and more recently an ISSI team International Space Science Institute (ISSI) on recalibration of the SN (https://www.issibern.ch/teams/sunspotnoser/).

One of the most stable sunspot databases was constructed by Professor Rudolf Wolf himself. In 1843, Professor Wolf founded a journal called the *"Mittheilungen der Naturforschenden Gesellschaft in Berne"* which continued as the *"Mitteilungen der Edgenossischen Sternwarte Zurich"*, when Wolf moved to Zurich in 1849. He published all of his findings annually in the latter collection. These findings include sunspot observations as far back as Harriot in 1610 (Wolf, 1861). The journal was maintained from 1848 (Wolf, 1848) until his death in 1893 (Wolf and Wolfer, 1894). He also collected sunspot records from his European colleagues and his auxilliary observers to tabulate them in these journals: in this article, they will be consistently referred to as *"the Mittheilungen"*.

The Royal Observatory of Belgium (https://www.astro.oma.be/en/), particularly the WDC-SILSO, conducted a mission between 2017 and 2019 to digitize all the data contained in the published *Mittheilungen*. Following these efforts, it is expected that we will be able to reconstruct the sunspot number series from the raw original data of individual observers. These raw data can be largely found in the *Mittheilungen*, but recent studies have partly revised and added individual sunspot and group observations on the basis of archival investigations from original materials (Carrasco *et al.*, 2020; Hayakawa *et al.*, 2020; Carrasco *et al.*, 2022, 2021; Carrasco, 2021; Hayakawa *et al.*, 2021).

Heinrich Schwabe was one of the most prominent contributing observers in the *Mittheilungen* with continuous solar observations from 1826 to 1868. As such, he is one of the "backbones" for Svalgaard and Schatten (2016); Clette *et al.* (2014) and Chatzistergos *et al.* (2017) reconstructions.

R. Wolf started his own observations from 1849 only and maintained the sunspot series in his scale from there. Even though Wolf only published his own observations during his Bern days, after moving to Zurich he started completing the gaps in his own sunspot observation series with that of Schwabe's (Friedli, 2016) (cf. Section 2). However, Wolf did not try to scale Schwabe's observations (Wolf, 1852) to his own until 1859, when he introduced the notion of calibration factors or k-factors. (Wolf, 1850d). Thereafter, Wolf used these k-factors (Clette *et al.*, 2007; Mathieu *et al.*, 2019) to calibrate all other observers to his own observations a posteriori back to 1849.

Recently, studies have reported a problem in the International SN series in 1849 when compared to other datasets (Leussu *et al.*, 2013). Thus, a thorough investigation of the corrections applied to all observations before 1848, including Schwabe's, is necessary to pinpoint the origin of this inconsistency in the SN series which coincides with the primary observer transfer from Schwabe to Wolf.

Our overarching goal being the Sunspot Number reconstruction, the validation of the homogeneity of observers before the "Wolf period" is an important step towards the construction of a robust SN series with the available raw data. In this study we present a thorough investigation of the 1849 transition between Schwabe and Wolf (Leussu *et al.*, 2013; Clette *et al.*, 2016; Friedli, 2016) in the context of the k-factors applied during the first construction of the International SN series, by exploiting all available datasets. For ease of comparison, we present our primary datasets in section 2 and an analysis of the Schwabe data in section 3. Section 4 gives the exact reason behind the inconsistency in 1849 in the original SN series (Version 1 of SN or SNV1). Sections 5 and 6 give the analysis of k-factors of the Schwabe series from 1849-1868 and 1826-1848 respectively. In section 7 we analyse the impact of this problem in 1849 on the remainder of the SN series and we present our conclusions in section 8.

## 2. Data Sources

### 2.1. Samuel Heinrich Schwabe

Samuel Heinrich Schwabe was born in Dessau (N51°50′,E12°14′) on October 25, 1789 and started his sunspot observations at the age of almost 36 in 1825. The first drawing of the solar disk was made on November 05, 1825, even though his first records of sunspots were found on October 1, 1825 and the last drawing was made on December 29, 1867, while the last verbal report is from December 31, 1867. More on his life can be found in Arlt (2011). Schwabe used a number of different telescopes for his observations from 1826-1867, although most of the changes can be found between 1825 and 1830. The most obvious and important changes are most probably linked to the learning process of making drawings: his early drawings (1825-1830) are coarser than the later ones. A detailed report on his observations and materials can be found

in Arlt (2011).

For 43 years, on almost every clear day, Schwabe observed the Sun and recorded its spots trying to detect a possible new planet among them. He did not find any planet but noticed a regular variation in the number of sunspots and published his findings in a short article entitled "Solar Observations during 1843" (Schwabe, 1844): this phenomenon is now known as the 11-year solar cycle. Schwabe's 39 logbooks are preserved in the archives of the Royal Astronomical Society in London and are still in very good conditions (RAS MS Schwabe 1-39). The majority of his drawings were made with a high-quality Fraunhofer refractor of 3.5 feet focal length (Arlt, 2011). The drawings have been recently subjected to analyses to derive sunspot positions, sunspot group number, individual sunspot number, and umbral areas (Arlt *et al.*, 2013) and were used in multiple studies on the space climate and space weather (Leussu *et al.*, 2013; Hayakawa *et al.*, 2019).
In the following sections, we categorise Schwabe's sunspot observations in four different datasets, based on the convenience for comparing them with each other.

### 2.1.1. Mittheilungen-(1826-1848)

Even though Schwabe observed since 1825, his first sunspot record appears in the *Mittheilungen* on January 5, 1826. Starting in 1849, Schwabe sent his new raw numbers on a monthly basis to Wolf to include them in the Sunspot Number series (SNV1), while his earlier data (from 1826 to 1848) were studied later by Wolf. Therefore, the Sunspot Number series before 1849 was scaled to Schwabe's observations (Friedli, 2016) which appear in *Mittheilungen* X (Wolf, 1850b), p247 (published in 1859, late after their documentation). We call this data set from January 5, 1826 - December 20, 1847, "Schwabe's Counts Part 1" or "SCP1", henceforth.

### 2.1.2. Mittheilungen-(1849-1868)

After 1849 Wolf used Schwabe's observations only when he himself did not observe to fill the gaps in his own observations (Wolf, 1878b; catalogue entry Hs368:46). Wolf might have received Schwabe's daily observations but only the observations he used to construct SNV1 can be found in *the Mittheilungen*. Moreover, during the 1849-1859 period, the data tabulated in the *Mittheilungen* as Wolf's observations are in fact, a mix of observations from different observers, with no way to differentiate. The first mention of Schwabe after 1848, in the tables of the *Mittheilungen* is in 1859 as a marker - indicating Wolf was not the only observer.

We call the Schwabe data from the *Mittheilungen* from January 1849 to December 1867 "Schwabe's Counts Part 2" or "SCP2" from now on. The first part of SCP2 (1849-1855) appears in *Mittheilungen* I (Wolf, 1850a), published in 1856, and the rest appears each year in the following *Mittheilungen* (Wolf, 1850d).

### 2.1.3. Wolf's Source Books

Other than the *Mittheilungen*, Wolf's own handwritten records on loose (unbinded) pages were recovered at the ETH Library in Zürich in 2015. These are known as Wolf's

Source Books. Wolf first recorded his observations in these "Source Books" and only later sent them for typewriting in digitized format, i.e. for publication in the *Mittheilungen*. In fact some of the data Wolf recorded in his Source Books (Wolf, 1878b; catalogue entry Hs368:46) he did not use, and thus might not have had them printed in the *Mittheilungen*. The Source Books from 1849 to 1877 were digitized by Friedli (2016), and appear on the Wolf Society website (http://www.wolfinstitute.ch/data-tables.html) and enable to identify exactly the days when Schwabe's observations filled Wolf's observational gaps. Data from before 1849 exists but remains to be digitized. We call the Schwabe data from the Source Books from January 1, 1849 to December 15, 1868 "Schwabe Source Books" or "SSB".

### 2.1.4. Reconstruction by Arlt et al. (2013)

As mentioned, Schwabe's original drawings have been digitised in Arlt *et al.* (2013), which provides a detailed record of these drawings including sunspot positions and sunspot umbral areas. They analysed about 135000 sunspots on scans of Schwabe's sunspot drawings from November 5, 1825 to December 29, 1867. A comprehensive explanation on the analysis methods can be found in Arlt *et al.* (2013). Verbal reports of spotless days are also included here. Regular quality controls ensure the homogeneity of the series. In this study we use the latest updated version found at https://old.aip.de/Members/rarlt/sunspots/schwabe/schwabe_tiltangles_v1.0_20150812.txt/view. Hereafter, we refer to this series and the corresponding reference as "A2013".

## 2.2. Rudolf Wolf

### 2.2.1. Mittheilungen data

Rudolf Wolf observed sunspots between 1848 and 1893 but started reporting his data from 1849 onwards. Over the years, he used several telescopes of which we were able to find clues in the *Mittheilungen* (Friedli, 2016; Bhattacharya *et al.*, 2021). Among these telescopes, he used observations from his standard refractor with an aperture of 37 Parisian lines, a focus length of 48 Parisian inches, and eyepieces for 64- and 144-fold magnification (Wolf, 1848; Friedli, 2016), as the reference of the International Sunspot Series and calibrated all other observers to this telescope's observations. In this study, we refer to this very dataset when we mention "Wolf's observations" throughout the article. All these observations can be found tabulated in the *Mittheilungen*.

### 2.2.2. Wolf's Source Books

Of course, Wolf's own observations can also be found in the Source Books (cf. section 2.1.3).

## 3. Stability of Schwabe Data (1825-1830)

As explained in section 2, Schwabe changed his telescopes frequently between 1825 and 1830 and his drawing patterns evolved over this time period (Arlt, 2011). Therefore, it is necessary to do a "reliability test" on this part of the Schwabe data, to include them in the reconstruction. We look for a comparison observer that would enable to characterize Schwabe's defects in observing before 1830. Hence, the comparison observer should overlap 1830, before and after. From Chatzistergos *et al.* (2017) and personal correspondence from Dr. Hisashi Hayakawa, we provide a list of possible candidates whose data had been traced.

1. Lindener 1804-1827 : this dataset is being digitized by R. Artl (Arlt and Vaquero, 2020), but stops in 1827, i.e. too early to diagnose a trend in the Schwabe data.
2. Derflinger observations (1802-1824) (Hayakawa *et al.*, 2020) stop ins 1824, thus before Schwabe's observations.
3. Tevel observations were recounted and published by Carrasco (2021) and span 1816 to 1836, however, they are very sparse.
4. Prantner's observations span 1804-1844, but are sparse and only continuous around 1812-1817. They were published by Hayakawa *et al.* (2021).
5. Pastorff observed over 1819- 1833 fairly continuously (c.f section 7.1.1). Data are available in *the Mittheilungen*. His original drawings have also been recovered and will be digitised soon by Dr.Hisashi Hayakawa.
6. Gruithuisen sunspot drawings span 1817-1848. They have been scanned by Dr. Hisashi Hayakawa, but are not digitized yet.
7. Arago observed from 1822-1830. Even though he does not overlap the turning point of 1830, his data is also a possible source of comparison with Schwabe data. His data can be found in *the Mittheilungen* (c.f section 7.1.1).

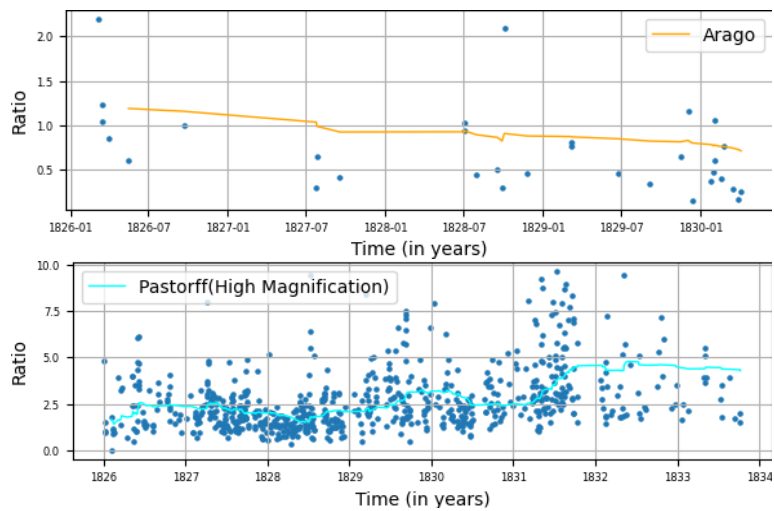

**Figure 1.** The panels shows the ratio between daily data of the Observer mentioned in the label/ Schwabe on overlapping days of observation. The solid line represents the moving average over 81 days (Mathieu *et al.*, 2019) and the dots represents the daily ratio. The x and y axes give the time (in years) and ratios respectively.

From the above list, and also from the list presented in section 7.1.1 we compare the daily data of all the observers with that of Schwabe's observations on overlapping days. Figure 1 shows this comparison for observers Arago and Pastorff, who have comparatively continuous data. We use the Mann-Kendall test to detect and quantify the trends on the daily ratios (Kendall, 1962). Arago shows a downward trend of about 19% whereas Pastorff shows an upward trend of about 22%. Tevel data from Carrasco (2021) also shows a downward trend of about 12% however, Tevel's data is too sparse for robust conclusions. Therefore, these comparisons are inconclusive because we cannot pinpoint the dataset responsible for the observed trends. The only way for a thorough study is the digitization of more key data from the above list that is scheduled, but not available yet. For now, we cannot exclude any data based on these inconclusive results.

## 4. Comparisons

Figure 2 (uppermost panel) shows the ratio of sunspot numbers (10*groups + spots, Wolf, 1848; Izenman, 1985) of A2013/SCP1 and A2013/SCP2. Here, SCP2 starts in 1849 thanks to the indications from SSB. We compared both datasets and found no difference, but used the indications of SSB to choose Schwabe data in the 1849-1859 period from the Mittheilungen.

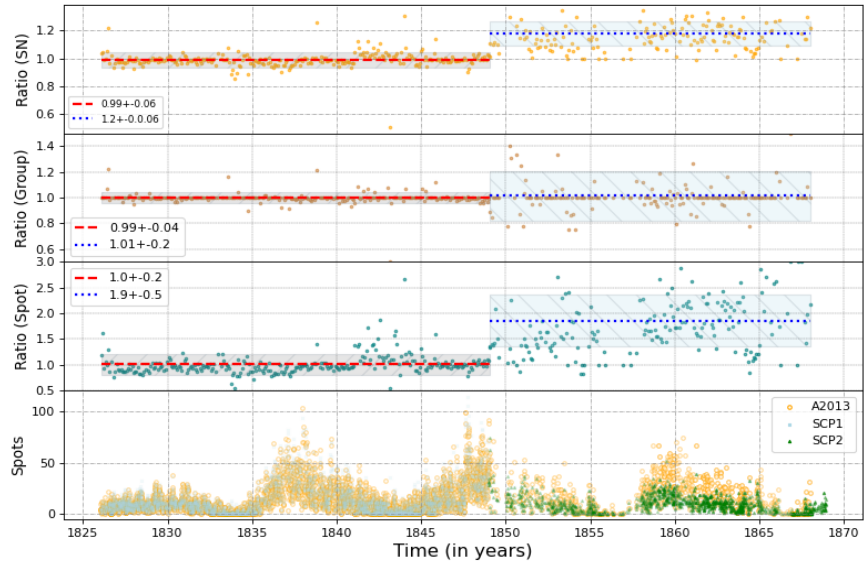

**Figure 2.** The upper panel shows the ratio of the monthly means of sunspot numbers A2013/SCP1 and A2013/SCP2 versus time. The dashed and dotted lines show the mean ratio of the two different timelines and the shaded regions corresponds to $1\sigma$ standard deviation. The corresponding panels show the same ratio of monthly means for the number of groups and spots respectively. The lower panel shows the daily spot counts in A2013, SCP1 and SCP2.

Figure 2 shows a change in the ratios with a level of about 22% even though all datasets come from the same observer, with its source either in A2013 or in SCP1 or SCP2. From Figure 2 (second and third panel), it also seems the group numbers approximately correspond between A2013 and the numbers present in the *Mittheilungen* for Schwabe (SCP1+SCP2). However, the number of spots differs by almost a factor of 2. This change of ratio in the number of spots takes place in 1849, long before 1859 when Schwabe's data became identifiable in the *Mittheilungen*. The lower panel of Figure 2 clearly shows that there is a downward effect in 1849 in the number of spots counted for Schwabe while the A2013 dataset does not show any noticeable change.

Note that we provide an overall mean of the 1849-1868 data even though the period from before 1859 seems to be slightly different. This is probably due to a combination of high solar variability with the scarcity of data (we do not have all of Schwabe data but only the used data, and thus, the volatility is very important) so we are concentrating on the overall jump value.

In addition to that, in the different studies that were made on A2013, i.e. Leussu *et al.* (2013), Senthamizh Pavai *et al.* (2015) and Bhattacharya *et al.* (2021), there is no event or change of behaviour between 1848 and 1849 that would account for such a huge difference. Thus we conclude that the difference comes from the *Mittheilungen* datasets, i.e the datasets for SCP1 and SCP2 are not the same, although they both come from Schwabe.

We investigate this change in the counts of groups and spots to pinpoint its origin.

In *Mittheilungen* X (Wolf, 1850b), (p247) Wolf writes: *Schwabe's sunspot observations in the years 1826 to 1848; The following communication contains observations, which I have made from the observation books of Mr. Heinrich Schwabe * concerning the years 1826 to 1848, following exactly the same principles which guided me in the earlier numbers when I communicated my own observations from the years 1849 to 1858, kindly supplemented by Mr. Schwabe.*

The above text is an English translation of the original German text found in the *Mittheilungen*. From this it seems that for the 1826-1848 period Wolf actually used the observation books from Schwabe and recounted the sunspots and groups himself, with the same method he used from 1849 to 1858 for his own observations. This implies the counts are close to his way of counting (correction factor k ≈ 1). However, from 1849 onwards, he got the data from Schwabe himself (Friedli, 2016) and, according to the Source Books he applied a correction factor of 1.25 (Wolf, 1878b; catalogue entry Hs368:46) calculated based on Schwabe's observations from 1849-1868.

A probable explanation for this difference in counting can be that Schwabe used to count one penumbra inside one spot whereas Wolf used the modern way of counting umbra inside spots. Figure 2 (third panel) shows the ratio of A2013/SCP2 is almost 1 near the minimum which indeed is evidence pointing towards this hypothesis. However more conclusive results should be obtained by comparing Schwabe's original drawings and countings.

To understand which correction factors were applied to the SCP1 and SCP2 datasets before including them in SNV1, we compute the ratio between SNV1 and the raw data from the *Mittheilungen* and Source Books as shown in Figure 3.

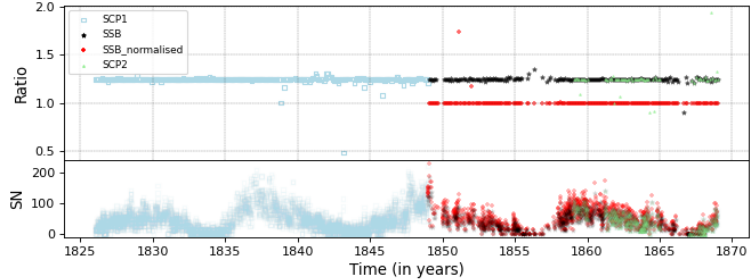

**Figure 3.** The upper panel shows the monthly ratio of SNV1 versus Schwabe *Mittheilungen* (blue squares), SNV1 versus source tables for raw Schwabe data (black stars) and normalized (x 1.25) Schwabe data (red diamonds). The lower panel shows the daily SN for SCP1, SSB, SCP2 and normalized SSB (mutliplied by its k factor) in the y-axis and time in x-axis

From Figure 3 it is evident that both SCP1 and SCP2+SSB have unfortunately been multiplied by 1.25 before inclusion into SNV1. Even though it is important to note that there was no direct way of comparing Wolf's observation with SCP1 (Schwabe pre-1849), the k-factors should not have been the same, as SCP1 and SCP2 are different datasets.

According to Figure 3 of Leussu *et al.* (2013) (our Figure 4), the ratio of SNV1/A2013 reflects a 21% jump (±2%). We now know that there should indeed be a jump in SNV1, of about 22% because Wolf wrongly applied 1.25 to the early Schwabe data, who was the primary observer from 1826-1848 (Friedli, 2016).

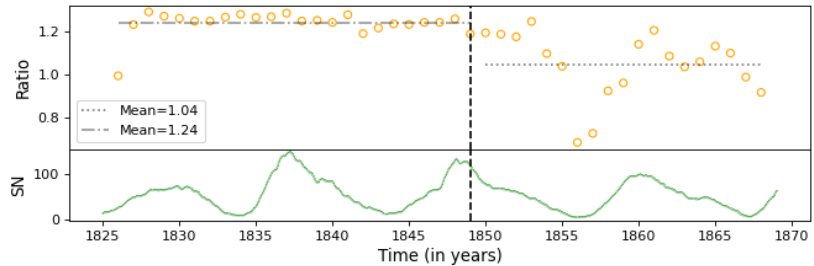

**Figure 4.** Yearly ratio for SNV1 versus A2013 with vertical broken line representing start of Wolf's observation as shown in Leussu *et al.* (2013).

This long-standing issue in 1849 in version 1 of the Sunspot Number series seems to have one origin : Wolf applied the same k factor to two datasets of Schwabe that were NOT counted in the same way. (1) The early data was recounted by Wolf himself in 1859 and (2) the later part was added "in real time" from Schwabe's correspondence. The later part (2) indeed showed a k-factor of 1.25 (cf. section 5) while the earlier part (1) should have a factor close to 1 that needs to be determined with precision using the datasets at our disposal. However, note that, at the time, Wolf did not have a viable comparison solution to compute another k-factor for the early part of Schwabe data. He should have realized his way of counting was different from Schwabe's way of counting (considering the 1.25 factor) but he did not have a reliable comparison source anyway.

## 5. Validation of SCP2 k-factor (1.25)

For verifying the k-factor applied to SCP2, we adopted various combinations of basic statistical quantification (mean and median) and we conclude the optimum k-factor that should have been applied to SCP2 is indeed $1.25 \pm 0.05$ as derived in Wolf (1862); Friedli (2016). Note that we resort to monthly averaged values as Wolf only tabulated Schwabe's observations on days when there was a gap in his own observations. Therefore, direct comparison of daily values are not realizable at this point.

### 5.1. Linear regression method

We fit a slope as shown in Figure 5, to the monthly smoothed SCP2 and Wolf observations using for accuracy, three methods: (1) the simple ordinary least square OLS(x,y), (2) the inverted OLS(y,x) and (3) a total least square fit TLS(x,y), taking into account that neither SCP2 nor Wolf's observation can be considered as a reference (Bhattacharya *et al.*, 2021). Monthly smoothing is realised using mean and median values as shown in Figures 5a and b, respectively. In other words Figure 5a gives the monthly mean-ed values of both the series mentioned in the x and y-axes, and Figure 5b gives the fit for monthly median-ed values of both the series, SCP2 and Wolf's observations. Note that the main difference between mean and median (panels a and b) resides in the robustness of the computation of the extremes: the median is more robust.

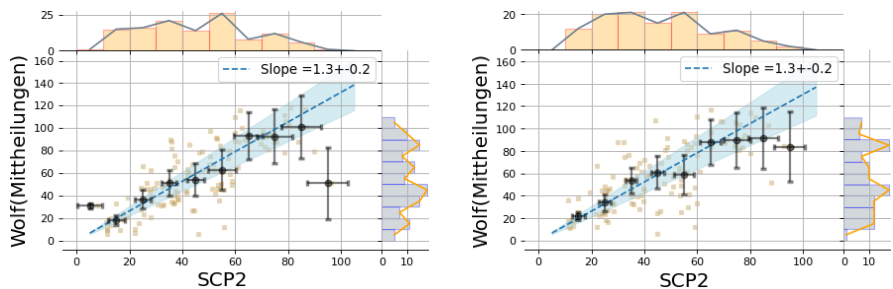

**Figure 5.** Regression using monthly mean(a) and median(b) values of SCP2 and Wolf observations using TLS(x,y). There are two panels in the figure (a) and (b). Each panel is consists of three panel the central, upper and lower. The central panel gives the mean monthly smoothed values in squares (a) and median monthly smoothed values in squares (b) with a regression line in blue with error as the shaded region and error bars in black. The upper panels in both (a) and (b) give the distribution of monthly smoothed values mentioned in the x-axis and the right panels give the distribution of the monthly smoothed values mentioned in the y-axis. The bin size is chosen as bin = 10.

The averaged regression coefficient using both fits as described above boils down to $1.3 \pm 0.2$, which is consistent with the 1.25 factor applied by Wolf.

## 6. Quantification of the SCP1-SCP2 jump using contemporary datasets

As explained in section 4, the application of 1.25 to SCP1 resulted in a jump around 1849. Although it is evident that SCP1 should not be scaled to Wolf's observations by a factor of 1.25, the determination of the actual scaling factor for SCP1 requires a different approach than the ones adopted in section 5 as there is no overlapping of SCP1 with Wolf data. In this section we use two different methods to investigate the "jump" with respect to several "reference series" overlapping the problematic transition, such as A2013 ors the group number series from Svalgaard and Schatten (2016) and Willamo, Usoskin, and Kovaltsov (2017).

### 6.1. Method Inspired by Lockwood, Owens, and Barnard (2014)

Because of the lack of overlap between the reference observers, we choose to use a method proposed by Lockwood, Owens, and Barnard (2014) for determining the "jump" value between SCP1 and SCP2. The authors employ a correction factor that should have been multiplied to one dataset (SCP1 here) to reach the level of the other (SCP2 here), such that:

$$\begin{aligned} \text{SCPc} &= \text{SCP2} : \text{for years} >= \text{jump year (1849)} \\ \text{SCPc} &= fc \times \text{SCP1} < \text{jump year (1849)} \end{aligned} \quad (1)$$

where SCPc indicates the whole series considered, i.e. the complete Schwabe dataset from 1826-1868. To compute this correction factor a stable reference series (Ref) is considered to compute residuals:

$$\text{Residuals} = \text{SCPc} - \text{Ref} \quad (2)$$

The next step of the methodology generates the averages of the residuals given by:

$$\begin{aligned} R_b &= < (\text{Residuals})_b > (\text{before}) \\ R_a &= < (\text{Residuals})_a > (\text{after}) \end{aligned} \quad (3)$$

$R_a$ and $R_b$ are then studied to determine the probability (Welch T-test) that $R_a$ and $R_b$ are the same, i.e $R_a$ - $R_b$ = 0. Welch T-test (WELCH, 1947) is equivalent to Student's T-test, except that it allows for both populations to have different variances and sample sizes, although their means ($R_a$ and $R_b$) are tested to be equals. A more comprehensive description of the above procedure can be found in Lockwood, Owens, and Barnard (2014).

We performed the method on our dataset using two different sets of estimators:

- A2013 and Group number constructed by Svalgaard and Schatten (2016) using backbone method (we call it BB16 from now on) for yearly averaged values.
- A2013 and the Group Number constructed by Willamo, Usoskin, and Kovaltsov (2017) using ADF method (we call U2016 from now) for daily values.

Both estimators can be considered as linear when compared with SNV1 although clearly not perfect. Note that we did not use the group number series constructed by Hoyt and Schatten (1998a,b) (HS98 from now on), due to its well-known inhomogeneity as mentioned in Cliver and Ling (2016), in the earlier part of nineteenth century. Also, we used several other estimators for which the results do not vary significantly, so they are not shown here.

Figure 6 a and b shows the results obtained for fc for reference group number series U2016 and BB16 respectively. Figure 6 a and b consists of two panels each, the upper panel shows the variation of $R_a - R_b$ as a function of fc which is varied from 0.9 to 1.5 at 0.0005 interval and the lower panel shows the variation of the probability density function of the Welsch T-Test as a function of fc. The optimum fc value is determined by the value of fc for which $R_a - R_b$ is 0 along with the maximum of the peak of the p.d.f plotted in the lower panels of Figure 6a and b. The dashed vertical line represents the weighted mean value of fc that are obtained for each individual reference viz. A2013, group numbers(BB16 or U2016) and estimator and the shaded region gives the $2\sigma$ uncertainty, that is 95% values lies within this region.

It is evident from Figure 6 (a and b), that the mean value of fc is $1.20 \pm 0.04$ and $1.19 \pm 0.03$ , considering $2\sigma$ values for the errors. Averaging the obtained values, the optimum fc value can be given as $1.2 \pm 0.05$ indicating there is a $20\% \pm 5\%$ jump between SCP1 and SCP2.

Note that as described in Section 2, Schwabe changed his telescopes frequently between 1825 and 1830 and his drawing patterns evolved considerably over this time period (Arlt, 2011). To check for a possible impact of this learning effect, we performed the above method excluding one year at a time from 1825 to 1835, i.e. on 1825-1848, then 1826-1848, etc... We find the data is mostly stable from 1830 onwards ($\approx 20\%$) whereas inclusion of 1825-1830 data gives the jump value 1% less than what was mentioned in Figures 6 (a and b). Nevertheless, all the values are within the errors provided. Figures 6 (a and b) represents the results using SCP1 data from 1830 to 1848 (i.e. excluding 1825-1829).

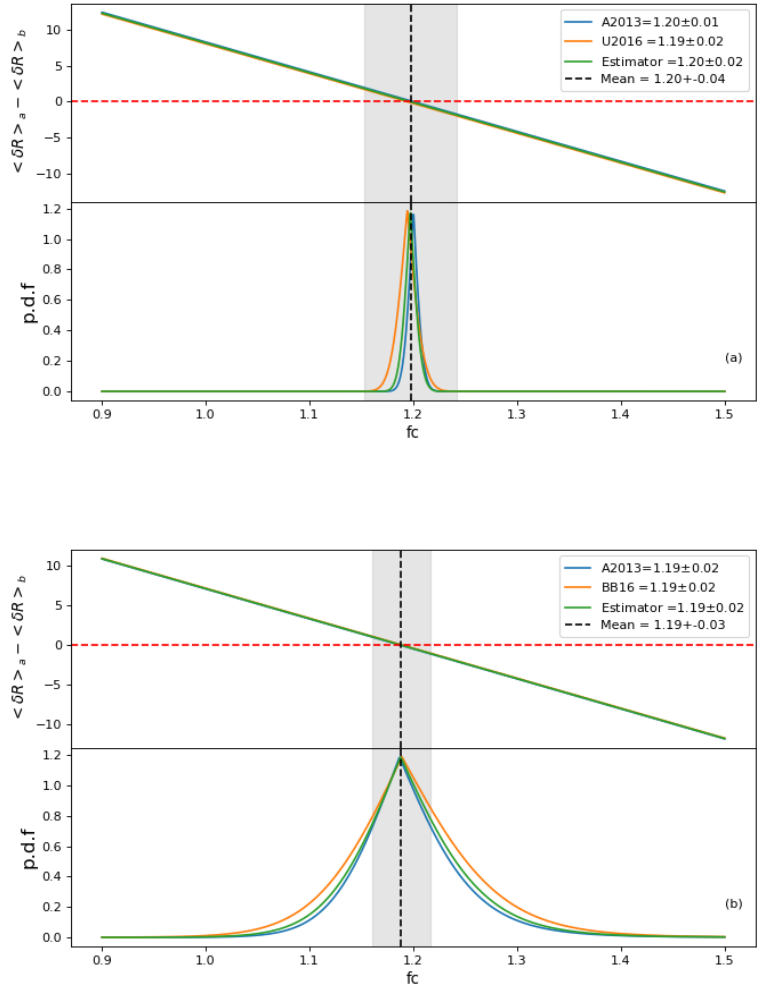

**Figure 6.** The upper panel (for both a and b) shows the variation of $R_a - R_b$ as a function of fc. The lower panel(for both a and b) shows the variation of the probability density function of the Welsch T-Test as a function of fc. The dashed vertical line represents the weighted mean value of fc that are obtained for each individual reference viz. A2013, group numbers(U2016(a) or BB16 (b)) and the estimator and the shaded region gives the weighted $2\sigma$ uncertainty.

### 6.2. Ideal k-factor of SCP1

As derived in section 5, the k-factor that should have been applied to SCP2 by Wolf, should be indeed around 1.25. From section 6.1 we determined that the application of 1.25 to SCP1 resulted in a jump of about $20\% \pm 5\%$.
Note that in the above approaches we assume A2013 to be a stable series. If we keep that assumption, we can calculate a simplified version of the k-factor "RX" that should have been applied to SCP1, below using basic ratios. The ratio of sunspot numbers of the following series can be expressed as R±DR where R is the ratio and DR is the associated error. Here "WOLF" refers to Rudolf Wolf's observations from period 1849 using his standard telescope, which is the reference series for SNV1.

Let us assume, WOLF/SCP1 = RX ± DX, A2013/SCP1 = R1 ± DR1, A2013/SCP2 = R2 ± DR2, WOLF/SCP2 = R3 ± DR3. Then RX = 1.03 ± 0.01. This means that Wolf should have multiplied the SCP1 dataset from Schwabe by 1.03 instead of 1.25, which means an error of 22 ± 1% on SNV1. This value is comparable to the values derived with methods described in section 6.1. Note that, since Wolf relied on calculations to derive the k-factors of the observers, we use the above derived value as the corrected k-factor for SCP1.

## 7. Backward and forward impact of the different SCP1 k-factor

After analysing all the results from section 6, it appears evident that the k-factor for SCP1 should be lowered by ≈ 22%. In this section we investigate whether lowering this k-factor impacts SNV1 before and after the SCP1 period (1825-1849) as well as Schwabe's contemporary observers who observed during the SCP1 period.

### 7.1. Backward Impact on Daily SN : 1818-1848

The daily sunspot number is available from 1818 to the present, hence we limit our investigation of the impact study for the daily sunspot number from 1818 only.

With the digitization of the *Mittheilungen* (digital format of Wolf's journals) it is possible to get access to all the raw data of all the observers involved in the construction of the SNV1 series (http://www.sidc.be/silso/versionarchive). However, there is no clear distinction in the *Mittheilungen* tables which were the exact data of each observer being used for a particular day.

#### 7.1.1. Identified Observers

Figure 7 gives the observers originally used by Wolf in the computation of SNV1 for the period 1818-1848. It is evident from this figure that from 1833 onwards, SNV1 is essentially Schwabe data (with an inaccurate k-factor of 1.25). Below, we develop the characteristics of the different observers from the above-mentioned period.

C. Tevel observed from 1816 to 1835. Wolf got his observations from Professor Buys-Ballot whom he met while visiting Academy of Sciences in Amsterdam (N52°22′,E4°53′). Tevel, "an astronomy lover", detected the first sunspot on April 2, 1816 and followed it for 2 days to see it's shape was changing and thereafter, from April 5, 1816 onwards he started a series of drawings using 125x magnification. Before his death he gave the drawings to "an association of learned men so that they may bring science to fruition", and after that the drawings were passed to Wolf by Professor Buys-Ballot. However, after 1820, there was a multi-year break until 1823 when he found no sunspots for the entire year and subsequently lost interest following the appearance of small groups. It was only after 1828 that he began his observations again after reading about a large sunspot group in the "Journal de Gand" (Wolf, 1850c). Recently Carrasco (2021) revisited Tevel's observations in detail.

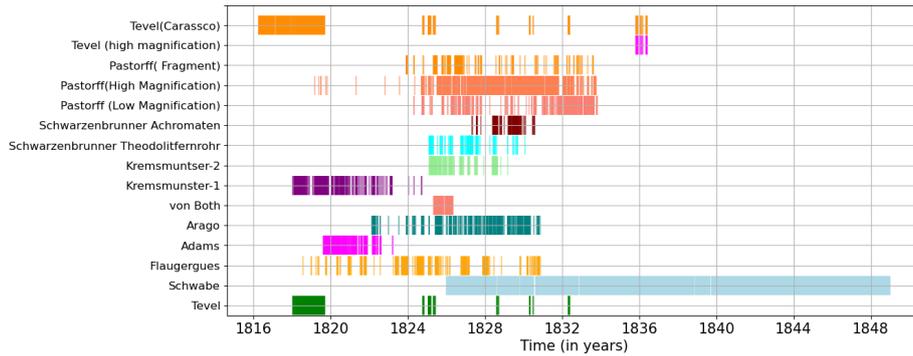

**Figure 7.** Observers available in the *Mittheilungen* along with the span of their observations during the period 1818-1848

Honoré Flaugergues of Viviers (N44°29′,E4°41′) was born on May 16, 1755 to a wealthy and educated family, and showed interest and talent for mathematics at an early age. He received an honorary mention as early as 1779 in Paris for a treatise "Sur la théorie des machines simples". He observed sunspots from 1788 till 1824 (Wolf, 1861). Wolf already mentions in the *Mittheilungen* that Flaugergues could be the link between Staudacher and Schwabe if he could find Flaugergues original diaries.

C.H Adams of Edmonton (N53°37′,W113°19′) observed the Sun between 1819 and 1823. His drawings were preserved in the Royal Astronomical Society and Richard Carrington presented his numbers to Wolf (Wolf, 1861).

Wolf found Francois Arago's observations from 1822-1830 in the Volume 2 of the "Mèmoires Scientifiques" (Wolf, 1862).

Lieutenant-General von Both in Breslau (N51°7′,E17°2′) gave his notebook to Wolf in which he entered a series of sunspot images with great diligence, adding notes for the years 1825 and 1826. It helped Wolf to complete the Sunspot Series by Flaugergues, Tevel, Arago, Schwabe, etc.(Wolf, 1872).

The observations of Kremsmünster Observatory (N48°03′,E14°07′) was sent to Wolf by Professor Fr. Schwab with the note "Our old observation journals contain scattered notes on sunspots from the years 1802–1830". However, different instruments were used in the course of the entire observation period. From 1802-1824 was a Brander azimuthal quadrant with a 5.5 feet long telescope was used for observing the sunspots. Observations from this period is labelled as Kremsmünster-1 (Figure 7).
For the years 1825-1830, all the observations of Kremsmünster Observatory were all made by P. Bonifaz Schwarzenbrunner, but not always with the same telescope: observations of the telescope with a 12-inch Borda circle supplied by Reichenbach with a magnification of 70 is labelled as Kremsmünster-2 (Figure 7). He also used an 12-inch Theodolitfernrohr or a four-legged Achromat with a magnification of 55, next to or instead of the Kremsmünster-2 telescope. These observations are labelled as "Schwarzenbrunner Theodolitfernrohr" and "Schwarzenbrunner Achromaten" respectively (Figure 7, Wolf, 1894).

C. Tevel's observations for the years 1835-1836 required a higher magnification (see section 7.1.2, Wolf, 1850c), hence denoted separately.

Sunspot observations by J. W. Pastorff from 1819 to 1833 came to John Herschel from the Royal Astronomical Society and Mr. A. C. Ranyard gave them to Wolf on his request. However, Pastorff sometimes used only a picture of the fragment of the sun for counting spots, as well as some observations needed a higher magnification (see Figure 7, Table 1, Wolf, 1874). These observations are indicated by stickers in the *Mittheilungen*.

Most of the presented details of the observers are translations of the German prologues written by Wolf in their respective *Mittheilungen* before introducing their observations.

Note that while most of the observers (Pastroff, Tevel, Schwarzenbrunner, Kremsmünster-2, Von Both, Arago, Flaugergues) have a direct overlap with SCP1, a few others (Kremsmünster-1, Adams) do not have any common days of observations.

### 7.1.2. k-factor Corrections

To assess whether the above-mentioned observers were scaled to SCP1, we perform a preliminary investigation of the prologues written by Wolf in the *Mittheilungen* before introducing any observer's data, which gave us valuable inputs about one of the important observer's k-factors between 1818-1836, (Wolf, 1850c): C.Tevel. A rough translation of the same is given as follows.

*" The comparison of these last observations by Tevel with those of Schwabe at the same time yields, for 20 days in the years 1828, 1830 and 1832:*
*the average group number:         5.0 Schwabe = 6.4 Tevel*
*relative number:         61.6 Schwabe = 80.7 Tevel*
*On the other hand, observations covered for 15 days from the years 1835 and 1836,*
*the average group number:         6.1 Schwabe = 4,2 Tevel*
*and relative number:         86.4 Schwabe = 61, 9 Tevel*
*so that the relationship has almost reversed, and it must therefore be multiplied by 1.83, if one wanted to regard Tevel as a constant observer, according to Schwabe, the relative numbers calculated for the years before the minimum of 1833 to make them similar to those calculated for the years after the minimum."*

From the above excerpt, Tevel's k factors before and after 1833 can be summarised as:

61.6/80.7 = 0.76 (before 1833)        with respect to Schwabe
86.4/61.9 = 1.40 (after 1833)         with respect to Schwabe

In addition to that, Wolf indicates that Tevel data from before 1833 (let us call them Tevel1) must be multiplied by 1.83 to reach Tevel data from after 1833 (that we will call Tevel2).

To identify the actual k-factors applied to each of the observers to include them in SNV1, we calculate the ratio SN/*Mittheilungen* Data for any particular observer, and subsequently calculate the mode of this specific ratio which is treated as the most probable k-factor for that observer.

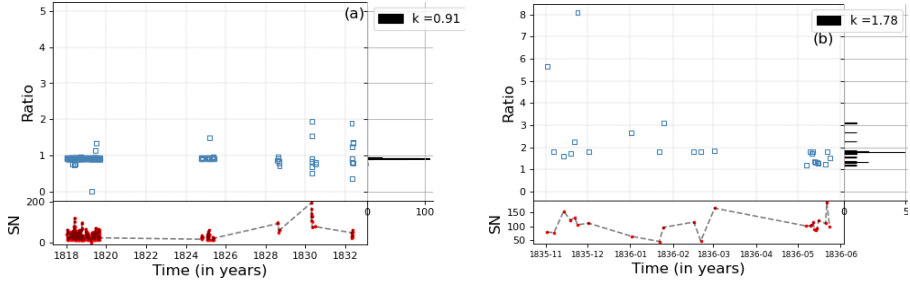

**Figure 8.** The left figure (a) is for Tevel's observation from 1818-1832 . The right figure (a) represents Tevel's observation between 1835 and 1836. The blue squares in both figures represent the daily ratio between SNV1/ Tevel. The lower panel gives the daily SN of Tevel's observation days. The right panel of both figures show the distribution of the ratio SN/Tevel, with the maximum frequency or mode indicated in the legend.

It is obvious from Figure 8 that the (most probable) k-factor for Tevel before 1833 is 0.91 and that for after 1833 is 1.78. As explained in the *Mittheilungen*, the k-factors for Tevel are: 0.76 and 1.40 respectively with respect to Schwabe (derived above).
It implies the applied k- factors are in fact, derived with respect to SNV1 which is constructed from Schwabe's observations after applying the incorrect k-factor of 1.25. That is:
(0.76 *1.25 = 0.95)         Applied : 0.91(before 1833)
( 1.40 * 1.25 = 1.74)        Applied: 1.78 (after 1833)
Hence, this clue points towards a backward influence of the erroneous application of 1.25 to SCP1 on the Tevel data.

We can also compute the corrected k-factor for Tevel using the error propagation method developed in section 6.2:

A2013/Tevel =R1 ± DR1 (cf. Figure 2), A2013/SCP1 = R2 ± DR2, WOLF/Tevel = X ± DX, WOLF/SCP1 = R3 ± DR3. Then Tevel/SCP1 = R2/R1 =R3/X

We compute the monthly smoothed ratio of the observers with A2013 (A2013/ Observers) with whom Schwabe had an overlap (R1) and their associated standard deviations (DR1). Therefore, RX(Tevel1) = 0.78 +/- 0.2. A similar value can be derived from the relative monthly sunspot numbers mentioned by Wolf himself in the *Mittheilungen*, i.e, the corrected k-factors for Tevel1 should be 0.76*1.03 = 0.78. For the values after 1833, RX(Tevel2) = 1.55 +/- 0.5 and the corrected k-factors for Tevel2 should 1.40*1.03 = 1.43. This implies that Wolf should have multiplied the dataset from Tevel1 by 0.78 instead of 0.91, and Tevel2 by 1.43 instead of 1.78 which indicates the jump of 1849 had a backward impact on all Tevel data.

The above calculations for Tevel1 and Tevel2 can be verified by a linear fit of Tevel and SCP1 as shown in Figure 9. The slope and errors are calculated in a similar way as explained in section 5.1.

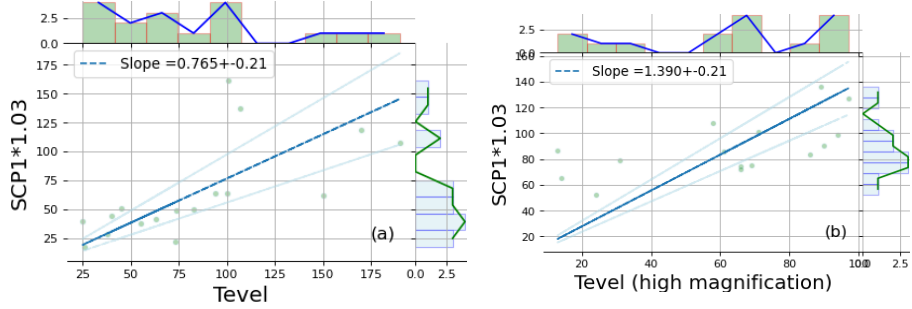

**Figure 9.** Regression using monthly means for Tevel's observations during 1818-1835(a) and Tevel's observations during 1835-1836(b) with the overlapping values of corrected SCP1 (SCP1*1.03 ± 0.02). The central panel gives the monthly smoothed values in blue circles(a),(b) with a regression line in blue with error as the shaded region. The upper panel gives the distribution of monthly smoothed values mentioned in the x-axis and the right panel gives the distribution of the monthly smoothed values mentioned in the y-axis.

We derive the applied and corrected k-factors of all the observers using the mode of the ratio SNV1/$SN_{observer}$ and determining the corrected k-factor by calculations similar to that of Tevel. We applied mode, as the ratios are mostly non-Guassian (see Figures 10a and b) and with given the skewness a robust estimator would be mode of the data. The summary of the applied and corrected k-factors of observers are listed in Table 1 columns 8 and 15 respectively. In Table 1 column 1 gives the ID of the observer as dedicated in the digitised *Mittheilungen*. Column 2 gives the corresponding name of the observer. Columns 3, 4, 5 gives the number of total observations available for the corresponding observer in the *Mittheilungen*, number of overlapped observations with Schwabe and number of zero observations respectively. Columns 6 and 7 gives the first day and last day of observation of the observer, respectively. Column 8 gives the applied k-factor derived by calculating SNV1/$SN_{Mittheilungen_{data}}$ as shown for Tevel in Figure 8. Columns 9-14 gives the coefficient values and their associated errors used for calculating the corrected k-factors for each observer, that is R1 = A2013/observer, R2 = A2013/SCP1, R3 = WOLF/SCP1 and the respective errors are denoted by DR1, DR2 and DR3. Columns 15, 16 and 17 gives the corrected k-factor (RX), associated standard deviation (DX) and the standard error (SE) respectively.Column 18 gives the ratio of the "Applied/X" k -factors. The "Applied" k- factor and "RX"- the corrected k-factors are colored in gray for easy comparisons.

Note that the information about the applied k-factors is only available for Tevel in the *Mittheilungen*. The k-factor derived for Tevel for the years after 1833 by direct calculation from *Mittheilungen* (1.43) is lower than its value given in Table 1 (1.55 ± 0.5). This ≈ 8% difference is due to the fact that there are just two years available for calculations after 1833, thus the error bar on this value is quite large (cf .Table 1). On the other hand, the applied k-factor for the period before 1833 is only ≈ 4% higher (it should have been 1.74 as derived above, according to calculations from *Mittheilungen* but applied 1.78). Nevertheless, all values are compatible within their error bars.

**Table 1.** Summary of the observers' data used to construct SNV1 from 1818-1849.

| FK_OBSERVERS | Name | No.of Obs | Overlapped obs | Zero obs | Start | End | Applied | R1 | DR1 | R2 | DR2 | R3 | DR3 | RX | DX | SE | Applied/X |
|---|---|---|---|---|---|---|---|---|---|---|---|---|---|---|---|---|
| 13 | Tevel | 355 | 20 | 15 | 1818-01-08 | 1832-05-18 | 0.91 | 0.75 | 0.22 | 0.99 | 0.06 | 1.03 | 0.01 | 0.78 | 0.23 | 0.01 | 1.17 |
| 22 | Flaugergues | 270 | 67 | 107 | 1819-01-01 | 1830-11-15 | 1.9 | 1.29 | 0.56 | 0.99 | 0.06 | 1.03 | 0.01 | 1.33 | 0.51 | 0.008 | 1.42 |
| 28 | Arago | 324 | 189 | 6 | 1822-02-15 | 1830-10-17 | 2.08 | 2.17 | 1.23 | 0.99 | 0.06 | 1.03 | 0.01 | 2.23 | 1.2 | 0.006 | 0.93 |
| 46 | von Both | 182 | 32 | 33 | 1825-05-04 | 1826-04-30 | 1.16 | 1.02 | 0.27 | 0.99 | 0.06 | 1.03 | 0.01 | 1.05 | 0.28 | 0.008 | 1.10 |
| 89 | Kremsmuntser-2 | 154 | 72 | 32 | 1825-02-08 | 1829-03-12 | 1.27 | 0.99 | 0.43 | 0.99 | 0.06 | 1.03 | 0.01 | 1.02 | 0.45 | 0.006 | 1.25 |
| 94 | Schwarzenbrunner Theodolitfernrohr | 110 | 63 | 17 | 1825-02-06 | 1830-01-26 | 1.27 | 1.04 | 0.39 | 0.99 | 0.06 | 1.03 | 0.01 | 1.07 | 0.4 | 0.006 | 1.19 |
| 96 | Schwarzenbrunner Achromaten | 135 | 103 | 1 | 1827-05-03 | 1830-07-27 | 1.14 | 0.97 | 0.33 | 0.99 | 0.06 | 1.03 | 0.01 | 1 | 0.3 | 0.003 | 1.14 |
| 131 | Tevel (high magnification) | 24 | 15 | 0 | 1835-11-02 | 1836-05-24 | 1.78 | 1.51 | 0.55 | 0.99 | 0.06 | 1.03 | 0.01 | 1.55 | 0.58 | 0.035 | 1.15 |
| 510 | Pastorff (low magnification) | 1155 | 759 | 0 | 1819-03-04 | 1833-10-12 | 0.46 | 0.45 | 0.13 | 0.99 | 0.06 | 1.03 | 0.01 | 0.46 | 0.12 | 0.0002 | 1 |
| 511 | Pastorff (high magnification) | 645 | 411 | 187 | 1823-12-02 | 1833-11-04 | 1.17 | 1.20 | 0.42 | 0.99 | 0.06 | 1.03 | 0.01 | 1.24 | 0.44 | 0.001 | 0.94 |
| 87 | Kremsmunster-1(Tevel) | 279 | 83 | 85 | 1818-01-13 | 1824-09-21 | 1.36 | 1.66 | 0.79 | 1.5 | 0.2 | 1.03 | 0.02 | 1.13 | 0.41 | 0.005 | 1.20 |
| 25 | Adams(Kremunster-1) | 402 | 88 | 94 | 1819-08-15 | 1823-03-16 | 1.33 | 0.90 | 0.64 | 0.6 | 0.3 | 0.78 | 0.2 | 1.17 | 0.54 | 0.006 | 1.14 |

As evident from Table 1, most of the observers found in the *Mittheilungen* need a modification in their respective k-factors as they were scaled to Schwabe (SCP1) from 1826-1849.
However, Arago and Pastorff's k-factors show a different trend. One of the possible reasons could be that Wolf chose too few of their observations to determine their k-factors when including them in SNV1. The scattering in Figures 10 (a) and (b) proves that indeed these observers require a re-evaluation of their k-factors if all of their data is to be included in the SN reconstruction. No further information regarding the k-factors of these observers can be found in the *Mittheilungen*.

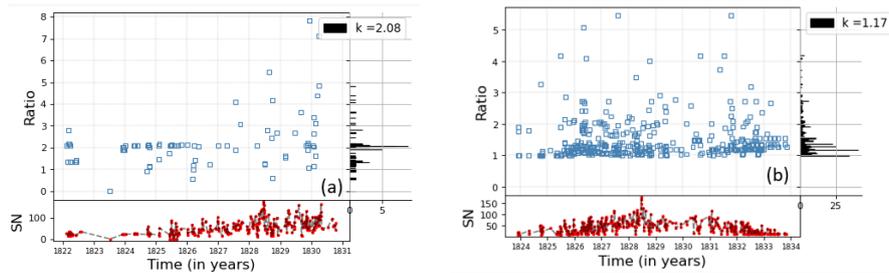

**Figure 10.** The left figure (a) is for Arago's Wolf Numbers. The right figure represents Pastorff (high magnification) Wolf Numbers. The blue squares in both figures represent the daily ratio between SNV1/ observer (Arago (a) or Pastorff (b)). The lower panel in both (a) and (b) gives the daily SN of the observers observation days. The right panel of both figures (a) and (b) show the distribution of the ratio SN/observer (Arago (a) or Pastorff (b)), with the maximum frequency or mode indicated in the legend.

Note that for the un-overlapped observers (Adams and Kremsmünster-1) with Schwabe, we use Tevel for determining the k-factor of Kremsmünster-1 and Kremsmünster-1 as reference series for Adams as mentioned in last two rows of Table 1.

### 7.1.3. SNV1 construction: identifying the observers

Identifying the observer for each day is a crucial step for implementing the corrected k-factors for each observer, to recreate a correct version of SNV1 (without the jump of 1849). Our adopted methodology for identifying the associated observer for a particular day data in SNV1 involves the following algorithm:

- Calculation of the ratio SNV1/$SN_{Mittheilungen_{data}}$ for any particular observer, and subsequently calculation of the mode (mo) of the ratio which is treated as the most probable k-factor factor for that observer (as explained in section 7.1.2).
- Diff = SNV1 - (Data * mo) is calculated.
- For each day we calculated sigma = $\sqrt{}$(SN) : considering SN to be Poisson (Dudok de Wit, Lefèvre, and Clette, 2016).
- Calculation of the max difference allowed = deviation of 3*sigma with respect to the day's SN, therefore allowed difference = $3 * \sqrt{}$(SN).
- For a day in SNV1 flagged as 0 implying, this is the only and only possible data available for the day:
    - If Diff > max difference : data is flagged as 3 (unreliable data).
- If there is more than one data available in *Mittheilungen* for a single day :
    - lowest Diff < max Diff, flag 1 (data was used).
    - Rest of the data are flagged as 2 (data was not used for this particular day).
    - if the lowest Diff > max difference, flag 3 (the data not reliable).
- For SN=0 days flag 4 is assigned:
    - If SN=0 but the data is simply an extrapolation of the previous or next day data then it is flagged (4,7)
    - If SN=0, but there is neither data in the *Mittheilungen* nor extrapolated, the data is flagged as (4,6).
    - If SN=0, but there is unused non-zero data in the *Mittheilungen* for one of the observers,the data is flagged as (4,2).
- If any day more than one data is present in the *Mittheilungen* and Diff gives the same value, flag 5 is assigned, implying both data gives same results.
- If there is a value in SN for a particular day and no value in *Mittheilungen*, flag 6 is assigned, implying this can be a erroneous value in the SN.
- For any data present in SN but not in *Mittheilungen*, but matches closely with either previous day data or next day data, it is flagged 7, indicating, this value in SN is an extrapolation.
- For days when there is no SN , they are flagged as -1.

An excerpt of this flagged table is given in table 1. The full data (1818-1848) can be found in https://www.sidc.be/silso/DATA/Schwabe-Wolf/Raw%20data%20of%20all%20the%20Observers%20(SN%20Version%201).csv. Note that there are only 23 flag 6 and 49 flag 3 data in the entire database from 1818 to 1848, validating our proposed methodology. In addition, we did not a fixed threshold as the maximum difference, as the deviation would be lower during low activity and extreme when SN is large.

**Table 2.** Excerpt of the table for pin-pointing exact raw data that are being used to construct SN version 1

| Date | FK_OBSERVERS | GROUPS | SUNSPOTS | WOLF | COMMENT | Sn | Mode | Name | Sn_R | Diff | Sigma | Threshold | flag |
|---|---|---|---|---|---|---|---|---|---|---|---|---|---|
| 01-02-1818 | NaN | NaN | NaN | NaN | NaN | -1 | NaN | NaN | NaN | NaN | | | -1 |
| 01-03-1818 | NaN | NaN | NaN | NaN | NaN | -1 | NaN | NaN | NaN | NaN | | | -1 |
| 01-04-1818 | NaN | NaN | NaN | NaN | NaN | -1 | NaN | NaN | NaN | NaN | | | -1 |
| 01-05-1818 | NaN | NaN | NaN | NaN | NaN | -1 | NaN | NaN | NaN | NaN | | | -1 |
| 01-06-1818 | NaN | NaN | NaN | NaN | NaN | -1 | NaN | NaN | NaN | NaN | | | -1 |
| 01-07-1818 | NaN | NaN | NaN | NaN | NaN | -1 | NaN | NaN | NaN | NaN | | | -1 |
| 01-08-1818 | 13 | 3 | 12 | 42 | Orfèvre à Middelburg (NL) | 39 | 0.91 | Tevel | 38.22 | 2 | 6.24 | 18.72 | 0 |
| 01-09-1818 | NaN | NaN | NaN | NaN | NaN | -1 | NaN | NaN | NaN | NaN | | | -1 |
| 01-10-1818 | NaN | NaN | NaN | NaN | NaN | -1 | NaN | NaN | NaN | NaN | | | -1 |
| 01-11-1818 | NaN | NaN | NaN | NaN | NaN | -1 | NaN | NaN | NaN | NaN | | | -1 |
| 01-12-1818 | NaN | NaN | NaN | NaN | NaN | -1 | NaN | NaN | NaN | NaN | | | -1 |
| 01-13-1818 | 87 | 1 | 6 | 16 | (NULL) | 22 | 1.36 | Kremsmunster-1 | 21.76 | 1.09 | 4.69 | 14.07 | 0 |
| 01-14-1818 | NaN | NaN | NaN | NaN | NaN | -1 | NaN | NaN | NaN | NaN | | | -1 |
| 01-15-1818 | NaN | NaN | NaN | NaN | NaN | -1 | NaN | NaN | NaN | NaN | | | -1 |
| 01-16-1818 | NaN | NaN | NaN | NaN | NaN | -1 | NaN | NaN | NaN | NaN | | | -1 |
| 01-17-1818 | 13 | 3 | 20 | 50 | Orfèvre à Middelburg (NL) | 46 | 0.91 | Tevel | 45.5 | 1.08 | 6.78 | 20.34 | 1 |
| 01-17-1818 | 87 | 1 | 2 | 12 | (NULL) | 46 | 1.36 | Kremsmunster-1 | 16.32 | 64.52 | 6.78 | 20.34 | 2 |
| 01-18-1818 | 13 | 4 | 24 | 64 | Orfèvre à Middelburg (NL) | 59 | 0.91 | Tevel | 58.24 | 1.28 | 7.68 | 23.04 | 0 |
| 01-19-1818 | 13 | 4 | 29 | 69 | Orfèvre à Middelburg (NL) | 63 | 0.91 | Tevel | 62.79 | 0.33 | 7.93 | 23.79 | 0 |
| 01-20-1818 | NaN | NaN | NaN | NaN | NaN | -1 | NaN | NaN | NaN | NaN | | | -1 |
| ... | ... | ... | ... | ... | ... | ... | ... | ... | ... | ... | ... | ... | ... |
| 11-02-1820 | 25 | 0 | 0 | 0 | X-25 to XI-7 nicht mehr bis | 0 | 1.33 | Adams | 0 | NaN | | | 4 |
| 11-03-1820 | 87 | 0 | 0 | 0 | (NULL) | 0 | 1.36 | Kremsmunster-1 | 0 | NaN | 0 | | 4 |
| 11-04-1820 | 25 | 0 | 0 | 0 | X-25 to XI-7 nicht mehr bis | 0 | 1.33 | Adams | 0 | NaN | | | 4 |
| 11-05-1820 | 25 | 0 | 0 | 0 | X-25 to XI-7 nicht mehr bis | 0 | 1.33 | Adams | 0 | NaN | | | 4 |
| 11-06-1820 | 25 | 0 | 0 | 0 | X-25 to XI-7 nicht mehr bis | 0 | 1.33 | Adams | 0 | NaN | | | 4 |
| 11-07-1820 | 25 | 0 | 0 | 0 | X-25 to XI-7 nicht mehr bis | 0 | 1.33 | Adams | 0 | NaN | | | 4 |
| 11-08-1820 | NaN | NaN | NaN | NaN | NaN | -1 | NaN | NaN | NaN | NaN | | | -1 |
| 11-09-1820 | 25 | 1 | 6 | 16 | (NULL) | 21 | 1.33 | Adams | 21.28 | 1.33 | 4.58 | 13.74 | 1 |
| 11-09-1820 | 87 | 1 | 1 | 11 | (NULL) | 21 | 1.36 | Kremsmunster-1 | 14.96 | 28.76 | 4.58 | 13.74 | 2 |
| 11-10-1820 | NaN | NaN | NaN | NaN | NaN | -1 | NaN | NaN | NaN | NaN | | | -1 |
| 11-11-1820 | NaN | NaN | NaN | NaN | NaN | -1 | NaN | NaN | NaN | NaN | | | -1 |
| ... | ... | ... | ... | ... | ... | ... | ... | ... | ... | ... | ... | ... | ... |
| 02-17-1834 | 17 | 2 | 3 | 23 | Extrapolation of 2-19 | 29 | 1.25 | Schwabe | 28.75 | NaN | | | 7 |
| 02-18-1834 | NaN | NaN | NaN | NaN | NaN | -1 | NaN | NaN | NaN | NaN | | | -1 |
| 02-19-1834 | 17 | 2 | 3 | 23 | (NULL) | 29 | 1.25 | Schwabe | 28.52 | 1.65 | 5.38 | 16.14 | 0 |
| 02-20-1834 | NaN | NaN | NaN | NaN | NaN | -1 | NaN | NaN | NaN | NaN | | | -1 |
| 02-21-1834 | 17 | 1 | 3 | 13 | Extrapolation of 2-23 day | 17 | 1.25 | Schwabe | 16.25 | NaN | 4.12 | 16.14 | 7 |
| 02-22-1834 | 17 | 1 | NaN | NaN | NaN | -1 | 1.25 | Schwabe | NaN | NaN | | | -1 |
| 02-23-1834 | 17 | 1 | 3 | 13 | (NULL) | 16 | 1.25 | Schwabe | 16.12 | 0.75 | 4 | 12 | 0 |
| 02-24-1834 | 17 | 1 | 1 | 11 | (NULL) | 14 | 1.25 | Schwabe | 13.64 | 2.57 | 3.74 | 11.22 | 0 |
| 02-25-1834 | 17 | 1 | 1 | 11 | (NULL) | 14 | 1.25 | Schwabe | 13.64 | 2.57 | 3.74 | 11.22 | 0 |
| 02-26-1834 | 17 | 0 | 0 | 0 | (NULL) | 0 | 1.25 | Schwabe | 0 | NaN | 0 | inf | 4 |

In Table 2 column 1 gives the Date, Column 2 gives the ID of the observer in digitised *Mittheilungen* database, Column 3,4 and 5 gives the groups, spots and Wolf Number of the observer as appears in the *Mittheilungen*. Column 6 lists comments specific to the flag assigned on that day. Column 7 lists the Sunspot Number Version 1 of the day. Column 8 and 9 gives the calculated k-factor (as explained in Step 1) of the observer and their respective names . Column 10 and 11 gives $Sn\_R$ = data*Mode and Diff = (SN - $Sn\_R$)/SN × 100. Column 12 and 13 gives the deviation sigma given by sigma = $\sqrt{SN}$ and allowed difference = 3*sigma (in percentage) respectively . Column 11 assigns flag to the data according to the above mentioned methodology.

This categorization enables us to pin point which observer's observation had been used each day for construction of SNV1, combined with the corrected k-factors we are now in a position to compute a more robust SNV1.

### 7.2. Forward Impact: 1849-1868

Prof. Rudolf Wolf started his observations from 1849 and thereafter, SNV1 has been maintained in his scale. In fact, the scaling factor of 1.25 for Schwabe was calculated based on SCP2 (Wolf, 1850a; Friedli, 2016), that is for the period (1849-1868) which overlaps with Wolf's observation period, and later applied inappropriately to SCP1. In this section we analyse the impact of the jump of 1849 on SNV1 from 1849 onwards. Note that we limit our forward impact analysis till 1868, to keep it within Schwabe's observation period (1826-1868).

#### 7.2.1. Identified Observers

Other than Schwabe, the observers used in SNV1 after 1849 and till 1868, are shown in Figure 11 and a summary of their details is listed in Table 3.

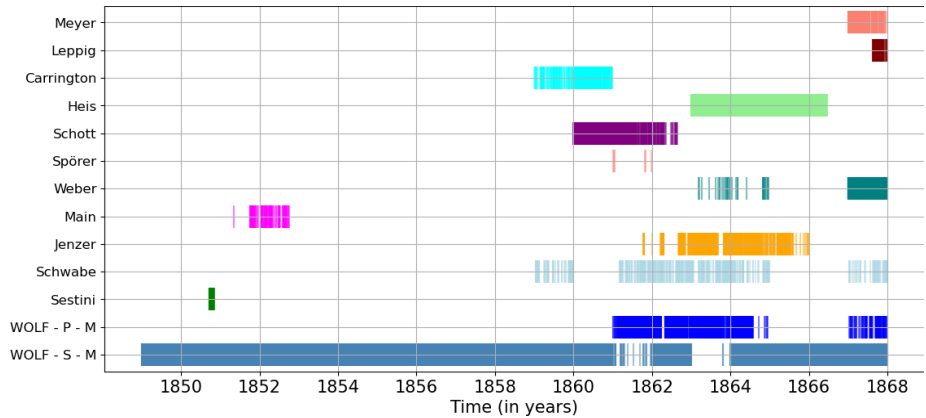

**Figure 11.** The observers available in the *Mittheilungen* along with the span of their observations from 1849-1868

Note that, Wolf's observations (WOLF-S-M) with standard telescope (see section 2.2.1) in the *Mittheilungen* from 1849-1859 is actually a combination of all the observers with no means to distinguish which observation is from which observer.

Wolf used his portable telescope (WOLF-P-M) or the Parisian refractor with 40 mm aperture and 700 mm focal length from 1852 until the end of 1889 and it served from 1861 as the main telescope for him and thus set the scale of the Wolf series for these years. It had a scaling factor of 1.5 with respect to Wolf's standard telescope (Wolf, 1875; Friedli, 2016).

Benedict Sestini was a Jesuit astronomer, mathematician and architect, who worked in Italy and the U.S. At Georgetown Observatory (N38°54′,W77°4′), in 1850, Sestini made a series of sunspot drawings, which were engraved and published (44 plates) as "Appendix A" of the Naval Observatory volume for 1847, printed in 1853. The work was republished in 1898 (Sestini, 1898).

Emil Jenzer was Wolf's pupil and observed from Bern (N46°56′, E7°26′) from 1861 to 1865 (Wolf, 1864; Friedli, 2016) and sported a scaling factor of 0.85 versus R. Wolf's observations. After Wolf's departure from Berne to Zurich Jenzer used the refractor which had an aperture of 83 mm and a focal length of 1300 mm.

Sir George Biddell Airy was an English mathematician and astronomer who was appointed Astronomer Royal in 1835 (Satterthwaite, 2014). Wolf received a letter from him on February 2, 1863 having sunspots observations from Greenwich Observatory (N41°1′, W73°37′) from 1851-1853 by Main having an k-factor of 1.5 and 25 other observations spanning from 1858-1860 from different observers for various purposes (Wolf, 1863).

Heinrich Weber observed from 1859-1883 in Peckeloh (N52°1′, E8°7′, 40km east of Münster, Germany)(Wolf, 1864).
Charles Anthony Schott was the superintendent of the Coast Survey (Washington D.C., US, N38°54′, W77°2′) and reported his observations to Wolf for the years 1860 and 1861. Wolf mentioned (Wolf, 1864) that the observations were made following the exact procedure introduced by Wolf, but they needed to be scaled by a factor of 1.16 to bring them to the same scale as his own. Wolf used these observations to fill gaps in his own observation for the years 1860 and 1861. Some of his original drawings can be found at https://www.ngdc.noaa.gov/stp/space-weather/solar-data/solar-imagery/photosphere/sunspot-drawings/charles-schott/.

Eudard Heis observed from 1863-1866 and again in 1872 from Münster (N51°57′,E7°37′) (Germany), and decided to share his observations with Prof. Wolf in 1863, as he observed in the same manner that Wolf introduced. (Wolf, 1864).

Richard Carrington's observations from his private observatory at Redhill (N51°14′, W0°11′) spans between 1853 to 1861.The English astronomer made a number of contributions to Solar Physics (Teague, 1996). His name is well-known because he was the first to observe a solar flare on 1 September 1859 which thereafter was renamed as the Carrington event (Cliver, 2006). He discovered the differential rotation of the Sun and created a heliographic coordinate system called the Carrington coordinates (Orchiston and Clark, 2007). However, Wolf only tabulated Carrington's spot counts for years 1859-1860 with a k-factor of 1.03 and reconstructed spot numbers from spot areas from years 1853-1858 (Bhattacharya *et al.*, 2021).

Gustav Adolf Meyer was Wolf's pupil who observed from 1867-1871 at Swiss Federal Observatory (Zürich, N47°22′, W0°34′) and sported a k-factor of 0.85 (Wolf, 1878a; Friedli, 2016).

Mr. H. Leppig reported observations at the Leipzig observatory (N51°20′, E12°22′) since 1867 August, partly with a 22-foot Dialyten and partly with a 4-foot Fraunhofer, using magnification 80 on both (Wolf, 1870).

In addition to the sources found in the *Mittheilungen*, Christian H. F. Peters observed the Sun from the Hamilton Observatory, in Clinton (N43°02′ ,W75°22′), New York (US) over the period 1860-1870 and his observations can be found in Casas and Vaquero (2014). Father Angelo Secchi also conducted sunspot observations in 1859 (Figure 4 of Hayakawa *et al.*, 2019), whereas his main observations are found in the 1870s in the *Mittheilungen* as well as in Vaquero *et al.* (2016).

### 7.2.2. k-factor Calculations

To determine any impact of the jump after 1849, we compute the k-factors applied to the observers after 1849. A monthly ratio (X) of the observer with that of Wolf's observations is shown in Table 3. Note that Wolf used a combination of observations using his standard telescope from 1849 and portable telescope from 1850, as mentioned in his source books (Wolf, 1878b; catalogue entry Hs368:46), therefore a normalised version of the combined observations is used for a stable and continuous coverage, that is, the Wolf's series use for calculating the ratios consists of:
StandardTelescope × 1 + PortableTelescope × 1.5 (Friedli, 2016; Wolf, 1878b; catalogue entry Hs368:46,).
However, the above mentioned Wolf series is from Wolf's source books (see section 2.1.3, Wolf, 1878b; catalogue entry Hs368:46), to avoid any biasing while calculating the k-factors as Wolf from *Mittheilungen* consists of all the observers from 1849-1859.

**Table 3.** Summary of the observers having observations in *Mittheilungen* after 1849.

| FK_OBSERVERS | Name | No.of Obs | Overlapped months | Zero obs | Start | End | Applied | X | DX | SE | Applied/X |
|---|---|---|---|---|---|---|---|---|---|---|---|
| 3 | WOLF - P - M | 933 | 57 | 62 | 1861-01-03 | 1867-12-23 | 1.5 | 1.5 | 0.066 | 0.001 | 1.00 |
| 16 | Sestini | 42 | 3 | 0 | 1850-09-19 | 1850-11-04 | 0.73 | 0.784 | 0.079 | 0.026 | 0.93 |
| 17 | Schwabe | 376 | 66 | 22 | 1859-01-13 | 1867-12-31 | 1.25 | 1.243 | 0.622 | 0.01 | 1.01 |
| 29 | Jenzer | 578 | 42 | 15 | 1861-10-14 | 1865-12-27 | 0.85 | 0.688 | 0.785 | 0.019 | 1.21 |
| 30 | Main | 98 | 14 | 1 | 1851-05-10 | 1852-10-02 | 1.5 | 1.516 | 0.863 | 0.045 | 1.00 |
| 32 | Weber | 387 | 27 | 214 | 1863-03-08 | 1867-12-31 | 0.73 | 0.729 | 0.454 | 0.018 | 1.00 |
| 33 | Spörer | 6 | 4 | 0 | 1861-01-07 | 1861-12-29 | 0.96 | 0.969 | 0.422 | 0.106 | 1.00 |
| 34 | Schott | 474 | 32 | 2 | 1860-01-01 | 1862-08-26 | 1.16 | 1.167 | 0.223 | 0.007 | 1.00 |
| 35 | Heis | 1107 | 40 | 74 | 1863-01-01 | 1866-06-18 | 0.78 | 0.748 | 0.392 | 0.01 | 1.04 |
| 36 | Carrington | 240 | 24 | 0 | 1859-01-02 | 1860-12-26 | 1.03 | 1.061 | 0.134 | 0.006 | 0.97 |
| 60 | Meyer | 201 | 13 | 124 | 1867-01-01 | 1868-01-01 | 0.85 | 0.685 | 2.173 | 0.198 | 1.14 |
| 42 | Leppig | 74 | 5 | 26 | 1867-08-19 | 1867-12-30 | 1.19 | 1.23 | 1.080 | 0.010 | 0.96 |

In Table 3 columns 1 and 2 give the ID and name of the observer as identified in the digitised *Mittheilungen*. Columns 3,4 and 5 give the total number of observations listed in the *Mittheilungen*, number of overlapped months with Wolf's observation and number of zero sunspot days in the whole data of the corresponding observer. Columns 6 and 7 gives the first and last day of observation of each observer. Columns 8,9,10,and 11 show the applied k-factor of each observer, the calculated k-factor given by Wolf/observer (monthly), associated standard deviation and standard error of each observer respectively. Column 12 gives the Applied/X ratio. The "Applied" k-factor and calculated "X" k-factor has been highlighted for comparability.

It is obvious from Table 3, the applied k-factors are within the error bars of the calculated k-factors and the scaling-factors indeed, had been applied with respect to Wolf's observations. Hence, we conclude that there is no significant impact of the 1849 jump on SNV1 after 1849.
For the observers who shows a deviation between the applied k-factors and our calculated k-factors, a thorough study needs to be performed with the Wolf's source book data, as much more information is available there on the observer's k-factors. For instance, Emil Jenzer (see section 7.2.1) was scaled using 3 different k-factors in the Source Books (http://www.wolfinstitute.ch/data-tables.html), 0.58 for 4 observations from October 31, 1862 to December 28, 1862, 0.49 for 7 observations from March 29, 1863 to Decembe 15, 1863, and 0.75 for 6 observations from January 14, 1864 to April 29, 1864. However, in the *Mittheilungen* his observations can be found from 1861-1865 with a k-factor of 0.85 (Wolf, 1862) and hence the deviation (see Table 3). Nevertheless we confine this study for the *Mittheilungen* data only.

We also omitted Schimdt, Franzenau and Kew Observatory from the forward impact analysis as Schmidt had only one overlapping observation for the considered period (1849-1868). Schmidt observed from 1841-1883 (Hoyt and Schatten, 1998b; Cliver and Ling, 2016) but in the *Mittheilungen* his observations were listed only from 1861 onwards. Franzenau observed from 1861-1863 but he used different instruments during the entire period without any distinction in his observations to identify which instrument had been used. As a result Wolf did not use his counts to fill the gaps (Wolf, 1864). The data from Kew Observatory had spot areas tabulated in the *Mittheilungen* instead of spot numbers for the years 1864,1865 and 1866 (Wolf, 1874), making them unsuitable for studying any impact of the jump.

### 7.3. Sunspot Number Series Version 2.1

As evident from Section 7.1, the existing SNV1 needs modifications in terms of the k-factors that had been applied to observers before including them in SNV1, before 1848. An obvious conclusion from Table 1 is that almost all the observers need a correction in their k-factors as they were all scaled to SCP1, but noticeably, not equally, hence, the entire SNV1 does not need to be lowered by the exact amount as SCP1 ($\approx 22\%$), before 1848.

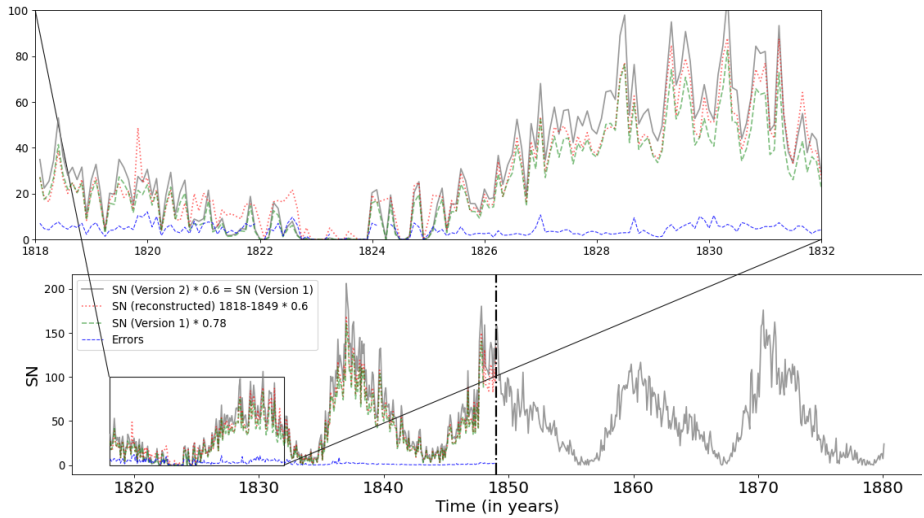

**Figure 12.** Monthly smoothed SNV1, SNV2, re-constructed SNV1 spanning from 1818-1868 and 22% lowered SNV1 from 1818-1848. The zoom in plot is for easy comparisons, with SN in the y-axis and time (in years) in the x-axis.

Figure 12 shows the monthly smoothed backward corrected SNV1, by applying the corrected k-factors to each observer for daily SN as explained in section 7.1.2, in red broken line from 1818-1849. The gray solid line signifies the existing monthly smoothed SNV1 and the blue solid line represents monthly smoothed SNV2. The green dashed line gives the monthly smoothed SNV1 after lowering it uniformly by 22% overall. The black broken vertical line represents the jump year 1849. The blue broken curve gives the daily errors given by $\delta\text{SNVx} = \sqrt{\sum_i((k_i * \delta\text{SN}_i)^2 + (\delta k_i * \text{SN}_i)^2)}$, where $k_i$ and $\delta k_i$ is the 'X' and 'DX' from Table 1 respectively, and $SN_i$ is the sunspot number for observer i for that particular day and $\delta SN_i$ is given by $\sqrt{(\text{SN}_i)}$, considering $SN_i$ to be Poisson (Dudok de Wit, Lefèvre, and Clette, 2016).

While the 22% lowered SNV1 and reconstructed SNV1 agree quite well after 1826, there is no significant difference between the reconstructed SNV1 and the existing SNV1 before 1826 as can be seen in the zoom-in plot of Figure 12. However, the scale seems to deviate again prior to 1823. Thus a simple lowering of 22% the SNV1 before 1848, is not enough and a reconstruction is necessary for that period.

Therefore, as an important step for the ongoing sunspot number re-calibration (Lefevre *et al.*, 2018), which resulted in SN Version 2 in July, 2015 (Clette *et al.*, 2016), we present SN Version 2.1 (https://www.sidc.be/silso/schwolftransition), in which SN Version 1 has been reconstructed from all the available raw data of the observers with revised k-factors, along with associated confidence bands as explained in section 7.1, for the period 1818-1848. We maintain the newly reconstructed version at the scale of SNV2. We do not find any significant scale discrepancies transmitted by that error in k-factor after 1849, hence no changes are included in SN (Version 2.1) after 1849. Even though it seems errors in SNV1 persist after 1849, they are caused by other factors than an error in k-factor or Schwabe data (e.g. observing techniques evolution, new observers, new telescopes, etc...).

The return of SNV1 to the same scale around 1826-1827 resonates with the idea put forward by Clette and Lefèvre (2016), where they find no propagation of the scale jump beyond the early Schwabe data (around 1830). They argue it could be attributed to Schwabe counting progressively more spots in the early part of his observations. Clette and Lefèvre (2016) limited their study till 1826 only, due to poor statistics of the available data prior to that period. Nevertheless, in this study, we go back till 1818, and by extending Figure 8 of Clette and Lefèvre (2016), a large deviation can indeed be found between BB16 and SNV1 as can be seen in Figure 13, in the shaded region.

Even though we find the SNV1 scale indeed deviates before 1823 (see the zoom-in plot of Figure 12), this large scale deviation is due the construction procedure of BB16. BB16 mentioned before the Schwabe backbone which started in 1830, they used the Staudacher backbone from 1749-1799. Flaugergues, who is an important bridge between Schwabe and Staudacher, is scaled to Staudacher (c.f Figure 21 of Svalgaard and Schatten, 2016). However, BB16 recounted Staudacher's observations themselves and reported almost 25% more groups than Wolf documented. As a result BB16 and SNV1 shows a large scale deviation prior to 1830.
The noteworthy conclusion is that the jump of 1849, which is clearly visible for SNV1/BB16, is eradicated with the newly reconstructed SNV2.1. The dip in the new series around 1832-1833 is due to the inclusion of Pastorff's observations in the new version which was omitted in SNV1.
The other comparison of SNV1 that Clette and Lefèvre (2016) reports is with Hoyt and Schatten (1998a) constructed GN (HS98; see section 6.1). HS98 involves a vast dataset including various observers which Wolf had not included in the *Mittheilungen* or in the SNV1 construction. Extending Figure 7 of Clette and Lefèvre (2016) backwards till 1818, we plot Figure 14. HS98 also depicts the jump of 1849, although the quantification of the jump is much lesser than the expected ≈ 22%. There is also an inexplicable upward trend which starts at around 1830 for SNV1 and SNV2.1). Since not much explanation is provided on how the k-factors are assigned to the observers during this period in HS98, this discrepancy remains unanswered. Nevertheless, our preliminary investigation through HS98 datasets and Cliver and Ling (2016), suggests an overall correction of their k-factors is necessary. For instance, as mentioned in section 7.1.2, Tevel was assigned distinct k-factors from 1816-1832 and 1833-1836 by Wolf. HS98, however, assigns a single calibration factor throughout the entire series in their GN construction. However, the 1849 jump is eradicated with the newly reconstructed series.

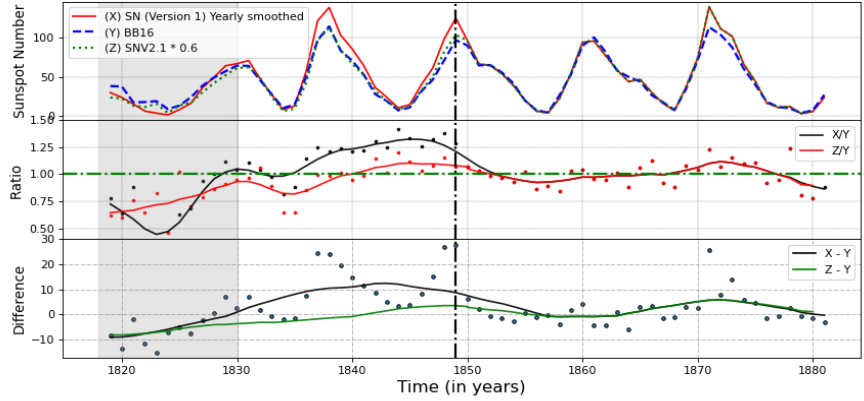

**Figure 13.** Comparison of the SNV1 and SN (Version 2.1) with the Backbone GN (BB16) over the years 1818-1880. The upper panel gives the original SNV1, SN (Version 2.1) and BB16 series. The middle panel gives the corresponding ratio and the lower panel gives the corresponding difference. The vertical broken line indicates the jump year 1849 and the shaded region gives the period when SNV1 and BB16 deviates largely.

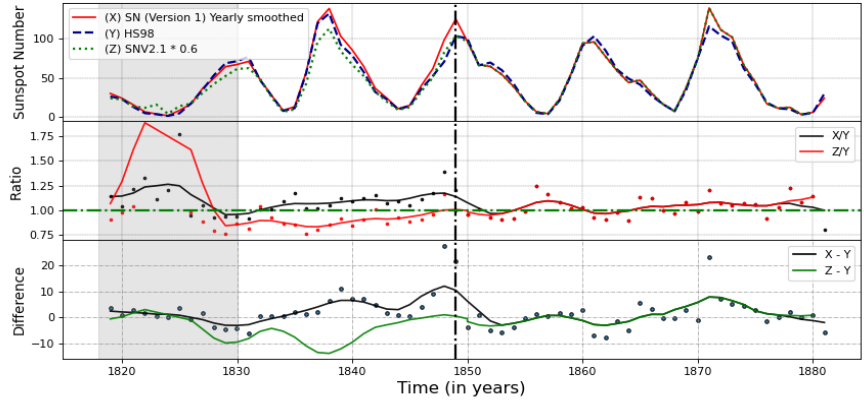

**Figure 14.** Comparison of the SNV1 and SN (Version 2.1) with the Hoyt and Schatten (1998a) constructed GN (HS98) over the years 1818-1880. The upper panel gives the original SNV1, SN (Version 2.1) and HS98 series. The middle panel gives the corresponding ratio and the lower panel gives the corresponding difference. The vertical broken line indicates the jump year 1849 and the shaded region gives the period when SNV1 and HS98 deviates largely. Only yearly values are plotted for consistency purposes.

Nonetheless, from the above Figures 12, 13 and 14, it is evident, that the k-factors Wolf applied to the early part of the Sunspot Series needs re-evaluations as he did not use enough observations to compute them. Even though, Clette and Lefèvre (2016) shows that the series did not necessarily transfer the jump, as it returns to the same scale as 1849, around 1830, but this phenomenon itself is a testimony that numerous mistakes in the application of k-factors remain. Thus, SNV1 indeed calls for a reconstruction from available raw data.

We limited the forward impact till 1868 to coincide with Schwabe's observation span (SCP2: 1849-1868, see section 7.2). Note that, we do not suggest any corrections on the SN after 1849 as we do not find any issues arising from Schwabe's data after the 1849 jump, as all observers were scaled to Wolf after 1849. In other words, the jump of 1849 specifically has not affected SNV1 afterwards. However, problems still exist which are not particularly linked to the use of Schwabe's data. For instance, it is evident from Figures 13, 14 and 15 and from the analysis by Clette and Lefèvre (2016) that the jump in 1848-1849 is resolved, but there remain lower artifacts (but still visible) from around 1868. The corrections proposed by (Clette *et al.*, 2016) applies until 1864 but the series comes back to the 1849 scale again in 1868. These need to be explored further by a similar analysis over the later part of the observations and the Wolf-Wolfer transition. Moreover, we intend to go beyond the notion of the k-factors and introduce methods for robust error determination on the SN series. That will be the subject of a subsequent study.

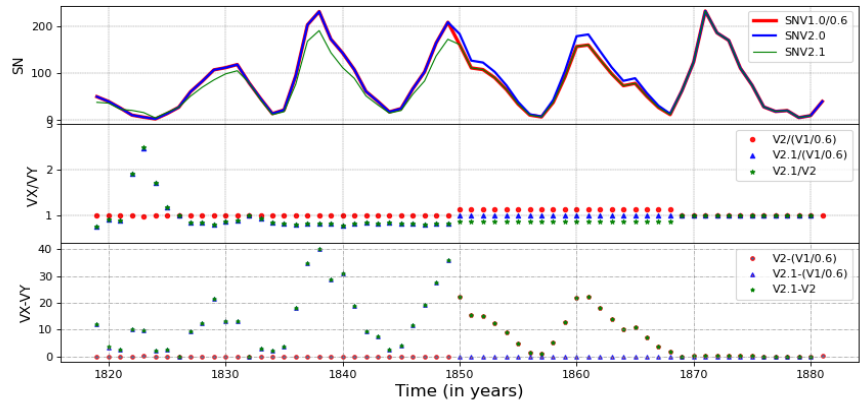

**Figure 15.** Comparison of the SNV1, SNV2 and SN (Version 2.1) 1818-1880. The upper panel gives the original SNV1, SNV2, SN (Version 2.1) . The middle panel gives the corresponding ratio and the lower panel gives the corresponding difference. Only yearly values are plotted for consistency purposes.

Note that, one could extract a Wolf's version of the reconstructed Sunspot Number series just as SNV1 would contain only Wolf's observer on a given day, by omitting Flag 2 data as explained in section 7.1.3, however Figure 12 shows the fully reconstructed series including all the available data of all the identified observers.

## 8. Conclusions

We present in this study the explanation for the ≈ 22%±1% jump in the International Sunspot Number series version 1 around 1849 which has been a long-standing issue and has been reported by several studies such as Leussu *et al.* (2013) and Friedli (2016). We propose the cause for this jump was due to the wrong application of the k-factor or scaling factor to Schwabe's observations before 1849. Wolf received Schwabe's observations as two separate series: one that he counted himself from Schwabe's drawings (from 1826-1848) and the later he received Schwabe's observations counted by Schwabe himself to fill the gaps in Wolf's observation series. However, the k-factor for Schwabe was calculated based on only the latter series and applied to the former as well, which resulted in the jump.

We assessed the stability of the early part of the Schwabe's data (1825-1830) through different comparison studies. The available data is not enough to prove any conclusive trend that might be present in the Schwabe early data. However, we pinpoint possible data sources that require digitization, which might help to detect any variation and correct it in later versions of the SN series. For the construction of SNV2.1 we did not exclude any data due to the lack of evidence on possible inhomogeneities.

We also validated the k-factor used by Wolf for Schwabe (1.25) from 1849-1868 by using various combinations of statistical quantities and linear regression. Since Wolf tabulated Schwabe's observation only on days he failed to observe, no daily comparison was possible, hence we resorted to monthly value comparisons. However, it is noteworthy that Wolf's journals *Mittheilungen* distinguishes Schwabe's observations after 1848, only in 1859. From 1849-1859, all observers are tabulated together in the *Mittheilungen* with no ways to distinguish on which day which observer has been used. Thanks to the Source books recovered and recently digitized by T. Friedli, it is possible to get a continuous series for Schwabe from 1849-1868, useful for validating the k-factor and thus, adding an error bar.

In addition to that, we quantified the jump using method inspired from Lockwood, Owens, and Barnard (2014). The methods are benefited from different overlapping series such as Group Number series by Svalgaard and Schatten (2016) and Willamo, Usoskin, and Kovaltsov (2017), and recounts of Schwabe's original observations by Arlt *et al.* (2013). After assimilating all the results, we propose a corrected k-factor of $1.03 \pm 0.01$ to Schwabe's observation before 1848.

We also assess the backward impact and forward impact of this jump on the International Sunspot Number series version 1. The daily sunspot number is available from 1818 to the present. A thorough investigation of the k-factors of the observers involved in the construction of the International Sunspot Number series version 1, revealed the jump has impacted the sunspot number series back to 1818, as from 1826-1848 Schwabe was the primary observer against whom all other observers were scaled. Hence the International Sunspot Number series needs to be revised with corrected k-factors of corresponding observers from 1818-1848. We also identified which observer has been used on which day in the International Sunspot Number series version 1 from 1818-1848 and flagged the raw data available in the *Mittheilungen* accordingly. We also present the corrected k-factors along with the associated errors for each identified observer, thus presenting International Sunspot Series version 2.1.
On the other hand, the application of an incorrect k-factor had no significant effect on the International Sunspot Number series version 1 from 1849 onwards.

We wish to remind the readers that the overarching goal of this work is the complete reconstruction of the Sunspot Number series from the raw data. Verification of the k-factors of the observers is an important step towards the reconstruction of the Sunspot Number series. A homogeneous Sunspot number series is necessary to study the evolution of long-term solar activity and its impact on the Earth (Carrasco, 2022). Moreover, the Sunspot Number series plays an active role in the accurate determination of the total solar irradiance models, thus, a reconstruction of the sunspot number with error bars is an absolute necessity for the scientific community at present (Lean, 2018).

We confined our study to the *Mittheilungen* data except for a few cases when we exploited the Source Books data but a thorough study is needed to include them in the reconstruction process as a lot more information is available, which the *Mittheilungen* lacks. As explained above, the reconstruction is here limited to the 1818-1848 period but will be extended in another paper to be published later.

**Acknowledgments** This work was supported by a PhD grant awarded by the Royal Observatory of Belgium to S. Bhattacharya. L. Lefèvre and F. Clette wish to acknowledge the support of ISSI https://www.issibern.ch/teams/sunspotnoser/.

**Disclosure of Potential Conflicts of Interest** The authors declare that they have no conflicts of interest.